\documentclass[twocolumn,showpacs,preprintnumbers,amsmath,amssymb]{revtex4}
\usepackage{graphicx}% Include figure files
\usepackage{dcolumn}% Align table columns on decimal point
\usepackage{bm}% bold math

\begin{document}

\title{Thermodynamics of Twisted DNA with Solvent Interaction}

\author{ Marco Zoli }
\affiliation{
School of Science and Technology - CNISM \\  Universit\`{a} di Camerino, I-62032 Camerino, Italy \\ marco.zoli@unicam.it}

\date{\today}

\begin{abstract}
The imaginary time path integral formalism is applied to a nonlinear Hamiltonian for a short fragment of heterogeneous DNA with a stabilizing solvent interaction term. Torsional effects are modeled by a twist angle between neighboring base pairs stacked along the molecule backbone. The base pair displacements are described by an ensemble of temperature dependent paths thus incorporating those fluctuational effects which shape the multisteps thermal denaturation. By summing over $\sim 10^7 - 10^8$ base pair paths, a large number of double helix configurations is taken into account consistently with the physical requirements of the model potential. The partition function is computed as a function of the twist. It is found that the equilibrium twist angle, peculiar of B-DNA at room temperature, yields the stablest helicoidal geometry against thermal disruption of the base pair hydrogen bonds. This result is corroborated by the computation of thermodynamical properties such as fractions of open base pairs and specific heat.
\end{abstract}


\maketitle

\section*{I. Introduction}

Nonlinearities in DNA dynamics were first emphasized  by Englander et al. \cite{englan} who interpreted the formation of temporary open segments of base pairs as mobile defects coherently propagating along the molecule backbone. That seminal paper put forward the idea that the open configuration, allowing for hydrogen exchange with the solvent, involved torsional oscillations of the base pairs around the backbone axis. Hence the thermally activated excitations of the double helix could be \emph{twisted} solitons spread over a length of about ten base pairs. Successive works \cite{proh} suggested that anharmonicity in hydrogen bond stretching modes could lead to self-trapping effects \cite{scott} and solitary-wave energy concentration thus providing the energy flow required in RNA transcription \cite{sawa}. Thus, soliton-like excitations were proposed to carry energy along the double helix in a way similar to the transport of biological energy occurring in $\alpha-$ helical protein molecules \cite{davy}.

Later on, it was pointed out \cite{pey1,pey2} that DNA biological functioning does not necessarily requires energy transport whereas nonlinearities should be rather considered in the framework of models able to overcome the continuum approximation thus accounting for the discreteness of the structure \cite{takeno,ctzhang,sale}.

A remarkable amount of studies published over the last decades \cite{wart,pey6}
has shed some light on the dynamics of DNA and the formation of fluctuational openings, the bubbles, which affect the thermal properties of the double helix including eventually its melting at high temperature.
To tackle these issues, two classes of theoretical methods have been developed. The \emph{first} is based on the Poland-Scheraga model \cite{poland,fisher} treating denaturing DNA essentially as a two state Ising-like chain of base pairs with regions of variable size, denaturated loops, opening temporarily due to thermal fluctuations.
The \emph{second} class assumes an Hamiltonian approach generally based on the Peyrard-Bishop model \cite{pey1} in which the transverse stretchings between complementary base pairs are represented by a one dimensional, continuous variable bringing the advantage that also intermediate states in the DNA dynamics can be described.
The Hamiltonian approach, describing the system at the level of the base pairs, seems particularly convenient in heterogeneous systems where the relative content of \emph{adenine-thymine}(AT) versus \emph{guanine-cytosine}(GC) pairs \cite{ares1} and sequence length \cite{theo1} are key to determine thermodynamics and melting of the double helix.

However, inside both theoretical approaches, different interpretations persist regarding the classification of the denaturation with some investigators finding a smooth crossover \cite{hanke,manghi1,santos1,theo2} whereas others point to a sharp transition driven \emph{either} by self-avoidance effects \cite{peliti,stella,carlon} \emph{or} by nonlinear stacking along the molecule backbone \cite{theo} \emph{or} by the finite range of the stacking itself \cite{joy05}. Denaturation in synthetic \emph{homopolymer} DNA displays a single melting temperature while \emph{heterogeneous } DNA fragments denaturate in multisteps between about $310$ and $410K$ depending on the sequence and salt concentration of the solvent \cite{cule,krueg}.

Microscopically, experiments based on proton-deuterium exchange methods \cite{gueron1} have shown that the base pair lifetimes are of order of milliseconds at $T \sim 305K$ with AT-pairs lifetimes being three times shorter than those of GC-pairs.
Fluorescence correlation spectroscopy \cite{bonnet} has further revealed that the breathing modes have a time scale in the $20 - 100$ microseconds range and bubbles of $2$ to $10$ base pairs open  at $T \sim 310K$ under low salt conditions.
Base pair openings are thus highly localized at biological temperatures and remain such also in homopolymers.

Theoretical modeling for DNA has then to incorporate nonlinear dynamics and large fluctuational effects which lead to base pair disruption and formation of bubbles whose size varies with temperature and environment \emph{pH} values \cite{metz06,metz07}. While fully atomistic descriptions become computationally very heavy even for short fragments, mesoscopic models have proven capable to capture the essentials of the interactions in DNA molecules. As fluctuations are expected to be strong in finite size sequences, I have recently proposed to apply the path integral formalism \cite{io,io1} to a one dimensional model, the Dauxois-Peyrard-Bishop (DPB) Hamiltonian \cite{pey2}, which incorporates nonlinearity also in the stacking potential. The idea underlying the method is that the base pair elongations with respect to the ground state are time dependent fluctuating paths which add to the partition function and cooperatively shape the denaturation transition in finite fragments.

Including in the computation a great number of molecule configurations, I have found that the denaturation is a smooth crossover both in homopolymers and heterogeneous systems. The latter also display the well known multistep melting features in the specific heat plots mirroring the fact that segments of the molecule open at different temperatures with AT-rich regions driving the bubble formation at relatively lower temperatures.
Interestingly a recent study \cite{singh}, based on the extended transfer matrix method \cite{zhang}, which focusses
on the same heterogeneous sequences considered in Ref.\cite{io1},  finds similar results regarding the temperature location of the main specific heat peaks.

However the Hamiltonian model proposed so far suffers of a serious shortcoming as it does not account for the helicoidal structure which would have the immediate effect of bringing closer to each other non-consecutive bases along the molecule backbone \cite{yakus1,gaeta}. As a first step to describe helicity in the path integral method one may generalize the stacking Hamiltonian by introducing the angle of rotation between a base pair and the previous one. In B-DNA at room temperature, one turn of the helix hosts about ten base pairs \cite{nelson}. Accordingly the equilibrium \emph{twist angle},  $\theta_{eq}=\, 2\pi /10.4$,  is expected to provide the energetically most favorable configuration, the stablest one against thermal disruption of the base pair bonds.

This Ansatz is a relevant benchmark for the computational method based on path integrals and should be viewed as a constraint for DNA theories. In this work  I develop the formalism for a generalized DPB Hamiltonian in which the twist angle between neighboring bases appears as a free parameter.
The Hamiltonian model also includes a solvent interaction term which realistically simulates the formation of hydrogen bonds with the solvent thus stabilizing the denaturated state \cite{druk}. The partition function has been computed by varying the strength of the solvent factors. In Section II the model Hamiltonian is presented while the main features of the path integral formalism are outlined in Section III. The thermodynamic results of the work are reported in Section IV where specifically the fractions of open base pairs, the specific heat and the base pair paths ensemble are calculated as a function of the twist angle. Some conclusions are drawn in Section V.

\section*{II. Hamiltonian Model}

The DPB Hamiltonian \cite{pey2} for a system of $N$ base pairs assumes the pair mates separation $y_n$ (for the \emph{n-th} base pair) with respect to
the ground state position  as the relevant degree of freedom. The longitudinal base pairs displacements along the molecule backbone are neglected as they are much smaller than the transverse stretchings $y_n$, hence the model Hamiltonian is essentially one-dimensional.  The  DPB Hamiltonian incorporates nonlinearities \emph{both} in the inter-base pair interactions modeled by a Morse potential \emph{and} in the coupling between neighboring bases along the two strands. As these are viewed as two parallel chains, the DPB model does not account for helicity.

\begin{figure}
\includegraphics[height=7.0cm,angle=0]{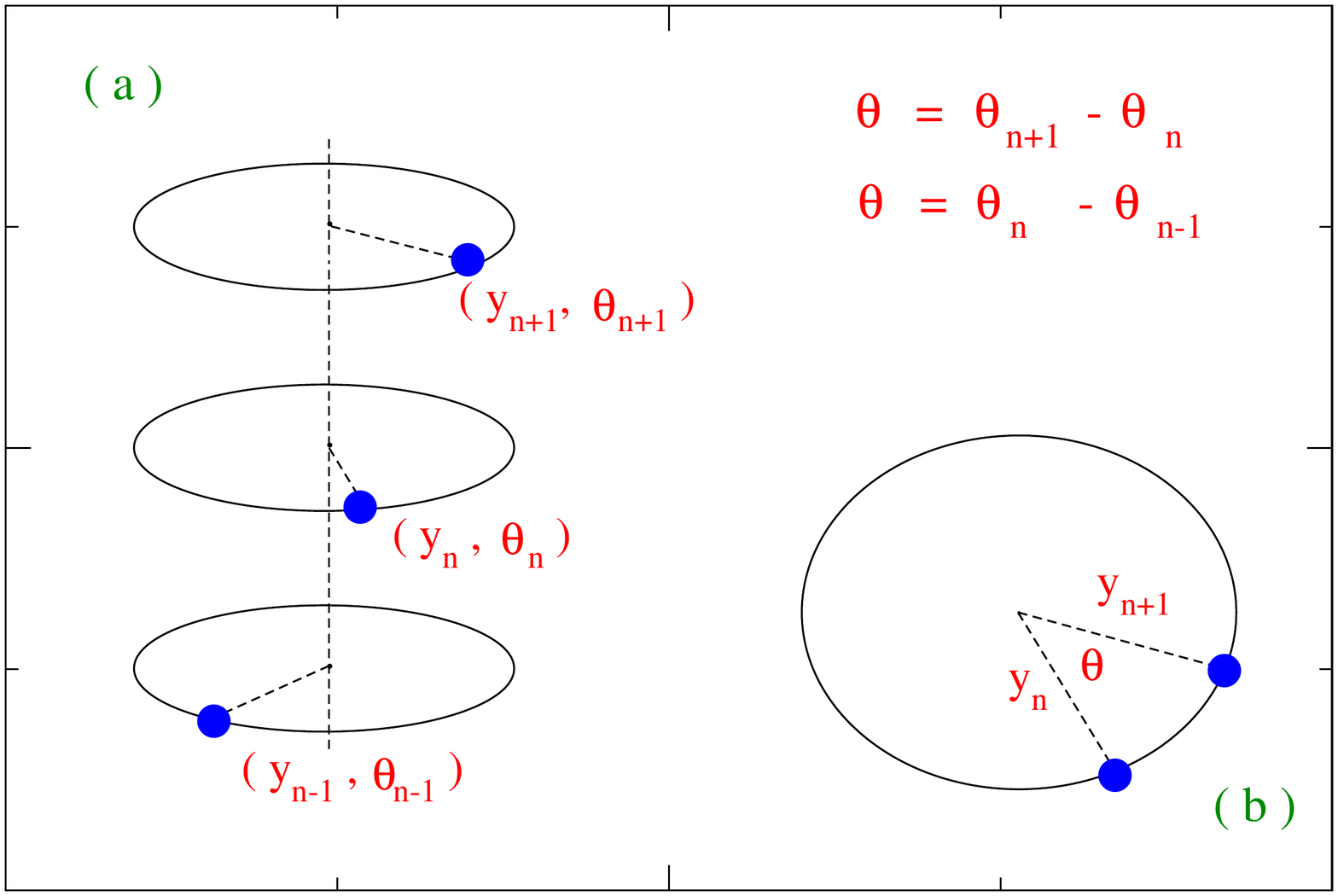}
\caption{\label{fig:0}(Color online) (a) Schematic fixed planes picture for the right-handed helicoidal model. The blue filled circles denote the pointlike base pairs stacked along the molecule axis with positive twist angle $\theta$. The radial coordinate $y_n$ describes the pair mates displacement in the $n-th$ base from the ground state. The dashed vertical axis corresponds to the $y_n \equiv 0$ configuration, the minimum for the one-coordinate potential $V_M(y_n) + V_{sol}(y_n)$ in Eq.~(\ref{eq:1}). There is no tilting between the planes.  (b) The helix plane seen from above.}
\end{figure}

Here, see Fig.~\ref{fig:0}, I introduce a twist angle $\theta$ between adjacent bases, $n$ and $n-1$, along the DNA backbone thus generalizing the DPB Hamiltonian to the following expression

\begin{eqnarray}
& & H =\, \sum_{n=1}^N \biggl[ {{\mu \dot{y}_{n}^2} \over {2}} +  V_S(y_n, y_{n-1}) + V_M(y_n) + V_{sol}(y_n) \biggr] \, \nonumber
\\
& & V_S(y_n, y_{n-1})=\, {K \over 2} g_{n,n-1}( y_n^2 - 2 y_n y_{n-1}\cos\theta + y_{n-1}^2 ) \, \nonumber
\\
& & g_{n,n-1}=\,1 + \rho \exp\bigl[-\alpha(y_n + y_{n-1})\bigr]\, \nonumber
\\
& & V_M(y_n) =\, D_n \bigl(\exp(-a_n y_n) - 1 \bigr)^2  \,
\, \nonumber
\\
& & V_{sol}(y_n) =\, - D_n f_s \bigl(\tanh(y_n/\l_s) - 1 \bigr)  \, .
\, \nonumber
\\
\label{eq:1}
\end{eqnarray}

$\mu$ is the base pair reduced mass which is related to the harmonic stacking coupling $K$ by $K=\,\mu \nu^2$, $\nu$ being the frequency of the phonon mode. As the DPB model is homogeneous with regard to the stacking interactions described by the potential $V_S(y_n, y_{n-1})$, $\mu$ takes the same value both for GC- and AT- bases. Also the anharmonic (positive) parameters $\rho$ and $\alpha$ are assumed independent of the type of base at the $n$ and $n-1$ sites and I have checked that such homogeneity assumptions don't change substantially the thermodynamical properties in the considered denaturation range. Instead, the latter are essentially determined by the size of $K$ \cite{yakov} as it will be shown in Section IV.A. Hereafter the values  $\rho=\,1$,  $\alpha=\,0.35 {\AA}^{-1}$ and $K=\, 60meV {\AA}^{-2}$ are taken consistently with previous investigations on the DPB Hamiltonian \cite{theo,io1,joy08}.  Different values for the stacking parameters are also found in the literature \cite{barbi1}.

Note that $\rho$ and $\alpha$ have a special role in driving the cooperative behavior of the system towards denaturation: when the molecule is closed, $y_n \,, \,y_{n-1} \ll \alpha^{-1}$
for all $n$, and the effective coupling is $K(1 + \rho)$. If, however, a fluctuation causes either $y_n > \alpha^{-1}$ or $y_{n-1} > \alpha^{-1}$, then the
hydrogen bond between pair mates loosens and the base moves out
of the strand axis. Accordingly the $\pi$ electrons overlap in the base plateaus is reduced, the binding between neighboring bases along the
strand weakens and the effective coupling drops to $K$ (in Eq.~(\ref{eq:1}), $g_{n,n-1} \simeq 1$). As a consequence, also the next base tends to move out of the stack thus propagating the fluctuational opening.
This explains the link between anharmonicity and cooperativity leading to bubble formation and eventually to denaturation in the DPB model.

The relevant feature is the torsion, due to the $\cos\theta$ term, in the stacking potential $V_S(y_n, y_{n-1})$ in Eq.~(\ref{eq:1}). While such torsional effects have been included in several mesoscopic studies, ranging from molecular dynamics simulations \cite{druk,campa1} to extensions of DPB model \cite{barbi0} and of Ising-like model \cite{manghi}, the role of the helicoidal geometry on the denaturation patterns is still unclear. To shed light on this issue I use the path integral method to study the thermodynamics of the system in Eq.~(\ref{eq:1}) as a function of $\theta$. Specifically a set of double helices is considered, each with $N$ base pairs whose arrangement along the strands is determined by $\theta$. The latter is taken as an input parameter. Only positive $\theta$ are considered to avoid non helicoidal structures with alternating $\pm \theta$ between consecutive base pairs along the molecule backbone: these zig-zag structures may also minimize the energy as $V_S(y_n, y_{n-1})$ is even in $\theta$.

Strictly speaking, the stacking varies with the torsion of the molecule thus the couplings should also bear a dependence on $\theta$, an effect which has not been quantified yet and it is here neglected.
It is emphasized that the model in Eq.~(\ref{eq:1}) is still one-dimensional as the rotational degree of freedom is site independent. Accordingly also the fluctuations around the equilibrium angle in the coiled structure are not accounted for. Moreover, as there is no tilting for the base pair planes in Fig.~\ref{fig:0}, bending effects \cite{marko} are not included in the model.

The base pair hydrogen bonds are modeled by the Morse potential in Eq.~(\ref{eq:1}): $V_M(y_n)$ incorporates heterogeneity in the sequence through the site dependent pair dissociation energy $D_n$ and the inverse length $a_n$ which sets the potential range. As the energy per hydrogen bond is about $15 meV$, I set for $AT-$ and $GC-$ bases respectively $D_{AT}=\,30 meV$ and $D_{GC}=\,45 meV$. AT-base pairs have larger displacements than GC-base pairs, then the inverse lengths are taken as $a_{AT}=\,4.2 {\AA}^{-1}$ and $a_{GC}=\,5 {\AA}^{-1}$  while slightly different values can be found in the literature \cite{campa,pey9}. Note that the plateau in $V_M(y_n)$ has a relevant physical implication: once all base pair displacements are larger than $a_n^{-1}$, the open strands can go infinitely apart with no energy cost. This means that the Morse potential does not account for strand recombination events which instead occur in solutions \cite{bonnet,pey3} and whose rate depends on the proton concentration in the solvent.  This drawback has been circumvented by several techniques which aim to restrict the configuration space thus keeping the transverse stretchings within a finite range \cite{zhang}. While the path integral method naturally operates a truncation of the path configuration space \cite{io1}, I treat here the problem analytically by adding in Eq.~(\ref{eq:1}) a solvent interaction term $V_{sol}(y_n)$ (as proposed in Refs.\cite{druk,collins}) modulated by a barrier factor $f_s$ and by a length $\l_s$ setting the range of the potential.

The solvent term has the effect to enhance (by $f_s D_n$) both the energy of the equilibrium configuration and the height of the barrier below which the base pair is closed. It is known that the melting temperatures depend logarithmically on the sodium concentration $[Na^+]$ in the solvent \cite{marmur,blake}. As the melting temperatures scale essentially linearly with the Morse potential barrier, also the latter can be assumed to vary logarithmically with $[Na^+]$ \cite{theo1}. This allows us to establish empirically a quantitative link between $f_s$ and $[Na^+]$. For instance, taking $f_s =\,0.1$,  one gets $[Na^+] \sim 0.4M$. Instead, the length $\l_s$ is taken so as to sample the effect of the solvent potential on the ensemble of the path displacements determined in the next Section.

\begin{figure}
\includegraphics[height=7.0cm,angle=0]{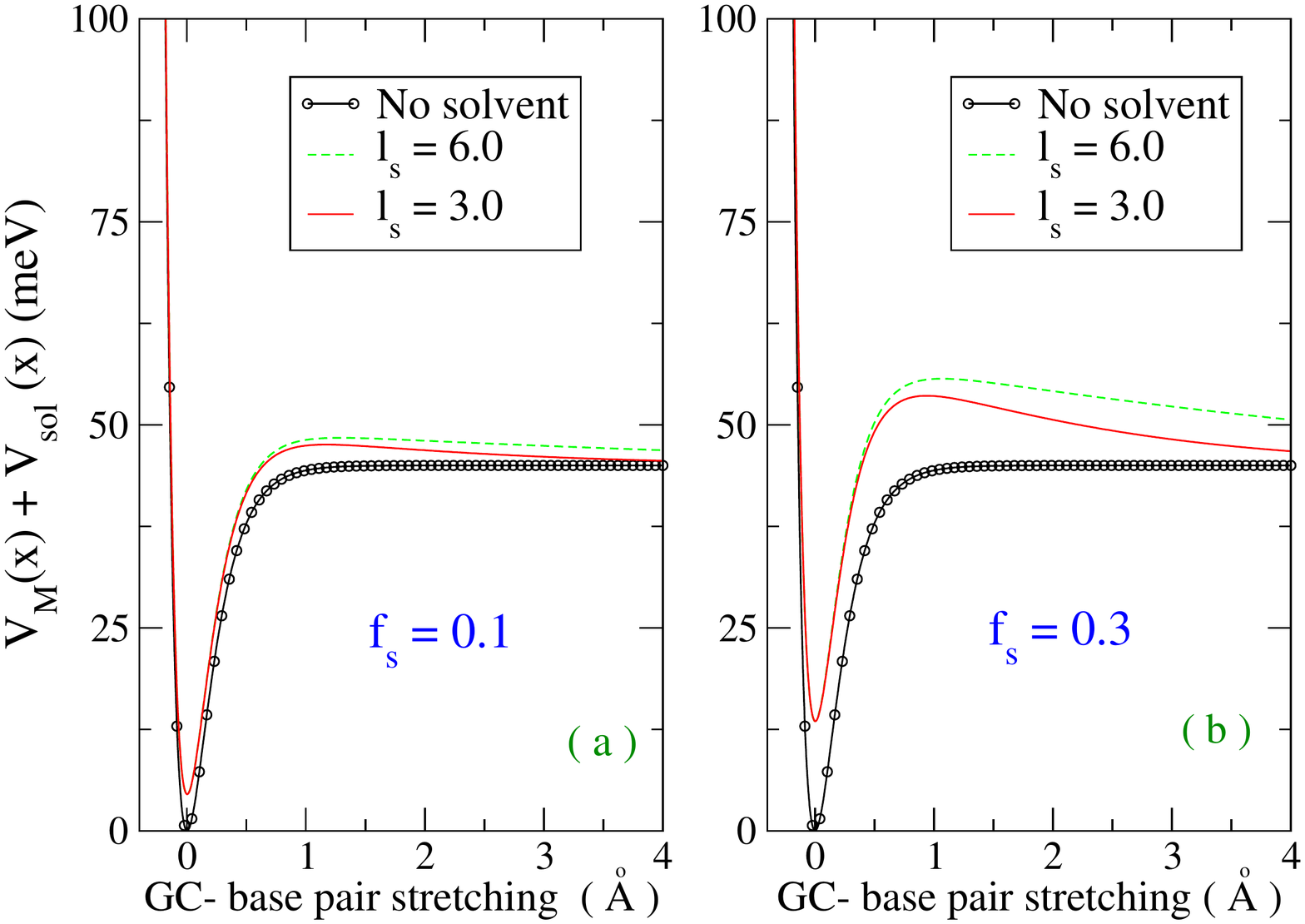}
\caption{\label{fig:1}(Color online) Sum of Morse $V_M$ and solvent $V_{sol}$ potentials versus \emph{guanine-cytosine} base pair separation for two values, in (a) and (b) respectively, of the solvent barrier factor $f_s$.  $l_s$ (in ${\AA}$) tunes the width of the solvent barrier. Also the bare Morse potential is plotted: the parameters $D_n$ and $a_n$ are taken for GC-base pairs. }
\end{figure}

In Fig.~\ref{fig:1}, $V_M(y_n)$ is plotted together with the superimposed effect due to $V_{sol}(y_n)$. The potential parameters are those for GC-base pairs and also the bare Morse potential is shown. As a main feature $V_M(y_n) + V_{sol}(y_n)$ displays a hump whose maximum determines the threshold stretching around which a base pair may first temporarily open and then either re-close or fully dissociate.
Looking at the case $l_s=\,3{\AA}$ in Fig.~\ref{fig:1}(b), the maximum occurs at $\sim 0.8 {\AA}$. Accordingly, in the range $0.8 < y_n < \l_s$, the pair mates are pulled away from each other and, for base pair stretchings larger than $\l_s$, the hydrogen bond with the solvent is established. In this sense the solvent interaction stabilizes the inter-strand open configurations.

\section*{III. Computational Method}

In the path integral formalism, the zero temperature evolution amplitude between two points, say "a" and "b", is a sum over {\it all} histories along which a system can evolve in going from "a" to "b" during the time $t$. Each history is weighed by a phase factor, the exponential of the action associated to a given path \cite{feyn}.
The formalism is extended to the finite temperature case, by performing an analytic continuation which defines the axis of the imaginary time $\tau$ with $\tau=\,it$. As $\tau \in [0, \beta]$ and $\beta$ is the inverse temperature, the imaginary time path integral permits to derive the thermal properties of the system by weighing the contributions to the action due to the particle paths $x(\tau)$ \cite{fehi}.

Accordingly the statistical partition function, which can be viewed as an analytic continuation of the quantum mechanical partition function, is given by an integral in the path phase space. Each path is weighed by a probability factor $exp(-A[x(\tau)])$ where the Euclidean action $A[x(\tau)]$ replaces the mechanical canonical action of the zero temperature case. Only closed paths contribute to the statistical partition function, the integration being a trace integration.
In the calculations, not {\it all} histories can be accounted for and, given the specific problem, one has to select the suitable class of paths which mainly contribute to the physical properties.

I have applied the path integral method to the discrete model in Eq.~(\ref{eq:1}) by introducing the idea that the $N$ base pair displacements $y_n$ can be described by one dimensional paths $x(\tau_i)$, the latter being periodic functions of the imaginary time $\tau_i$, $x(\tau_i)=\,x(\tau_i + \beta)$. The index $i$ numbers the base pairs along the $\tau$-axis. In fact, there are  $N + 1$ base pairs in Eq.~(\ref{eq:1}) but the presence of an extra base pair $y_0$ is remedied by taking periodic boundary conditions, $y_0 = \, y_N$, which close the finite chain into a loop.  This condition is  incorporated in the path integral description as the path is a closed trajectory, $x(0)=\,x(\beta)$. Hence a molecule configuration is given by $N_\tau \equiv N$ paths and, in the discrete imaginary time lattice, the separation between
nearest neighbors base pairs is $\Delta \tau =\,\beta / N_\tau$. Then, Eq.~(\ref{eq:1}) is mapped on the time axis as follows:

\begin{eqnarray}
& &y_n \rightarrow x(\tau_i) \,  \nonumber
\\
& &y_{n-1} \rightarrow x(\tau')\, \nonumber
\\
& &\tau' = \tau_i - \Delta \tau \,\nonumber
\\
& &n =\, 1\,,..\,,\,N  \, \,\nonumber
\\
& & i =\,1\,,..\,,\,N_\tau + 1 \, \nonumber
\\
& &\tau_1 \equiv 0 \,; \, \,\tau_{N_\tau + 1} \equiv \beta \,.
\label{eq:3}
\end{eqnarray}

Accordingly also the real time derivative $\dot{y}_{n}$ maps onto the imaginary time derivative $\dot{x}(\tau)$ as:

\begin{eqnarray}
{{d {y}_{n}} \over {dt}} \rightarrow (\nu \beta){{d x} \over {d\tau}}\,. \,
\label{eq:5}
\end{eqnarray}

Eq.~(\ref{eq:5}) is consistent with the replacement

\begin{eqnarray}
\hbar \rightarrow (\nu \beta)^{-1} \,,\,
\label{eq:6}
\end{eqnarray}

which is justified in the classical regime appropriate to DNA denaturation.
Eq.~(\ref{eq:6}) is also used to solve the pseudo-Schr\"{o}dinger equation for a Morse potential \cite{landau}  obtained from Eq.~(\ref{eq:1}) in the large $K$ limit (with $\rho=\,0$ and $f_s=\,0$) \cite{pey1}.

Due to their periodicity property the paths can be expanded in Fourier series with cutoff $M_F$

\begin{eqnarray}
& &x(\tau_i)=\, x_0 + \sum_{m=1}^{M_F}\Bigl[a_m \cos(\omega_m \tau_i) + b_m \sin(\omega_m \tau_i) \Bigr] \, \nonumber
\\ \,
& &\omega_m =\, {{2 m \pi} / {\beta}}\, \nonumber
\\
\label{eq:6a}
\end{eqnarray}

and this introduces the following physical picture:

\emph{a)} given a set of coefficients $\{x_0 , a_m , b_m\}$, the $N$ base pairs are represented by the configuration $\{x(\tau_i), \, i =\,1\,,..,\,N_\tau \}$.

\emph{b)} A set of coefficients corresponds to a point in the path configuration space thus, sampling the latter amounts to build an ensemble of distinct configurations for the system. As this is done for any temperature, we have a tool to describe the base pair thermal fluctuations around the equilibrium ($x(\tau_i) \sim 0$).

\emph{c)} In principle the configurations ensemble for the DNA fragment is infinite as it may include any possible combination of Fourier coefficients. For practical purposes some physical criteria intervene to select computationally the path coefficients defining a molecule configuration and contributing to the partition function. This poses a restriction on the ensemble size.

Such criteria are of two types: \emph{First}, the hydrogen bond potential in Fig.~\ref{fig:1} excludes too negative base pair stretchings due to the hard core which mimics the repulsion between negatively charged sugar-phosphate groups. Also too large displacements, $x(\tau_i) \gg \l_s$, are eliminated as they have no effect on the free energy derivatives. Once the pair mates are unbound and tied to the solvent molecule
there is no physical reason to further pull them to any particular direction. \emph{Second}, the ensemble of paths has to be consistent with the thermodynamics laws. This means that the numerical code selects, at any temperature, a path ensemble and evaluates the entropy of the DNA fragment. If the entropy is growing versus $T$, the code proceeds to the next temperature step otherwise a new partition is performed in the Fourier coefficients integration, a new path ensemble is selected and the entropy is recalculated. This is done at any $T$ until the macroscopic constraint of the second law of thermodynamics is fulfilled throughout the whole investigated temperature range. I emphasize that the method does not put any constraint on the shape of the \emph{entropy versus $T$}- plot aside from the requirement that the entropy derivative has to be positive.

It follows that the path ensemble is a dynamical object accounting for the manyfold of molecule configurations which enter the thermodynamical calculation. The size of the ensemble is a measure of the cooperativity degree of the system. By increasing $T$, some base pairs may open and cooperatively lead to bubble formation along segments of the double helix. Accordingly the ensemble size is expected to grow versus $T$.
I define $N_{eff}$ the number of sets of Fourier coefficients in Eq.~(\ref{eq:6a}), selected at a given temperature, which fulfill the criteria above. Then $N_{eff}$ is the number of possible trajectories for the $i-th$ base pair while the ensemble size for the whole fragment is measured by $N_\tau \times N_{eff}$. This is a key parameter in our analysis as it will be shown below.

So far, one molecule which may exist in $N_{eff}$ configurations has been assumed. In a picture closer to the experimental viewpoint, one may consider an ensemble of identical molecules (with $N_\tau$ base pairs). Each molecule may exist, at a given $T$, in a configuration specified by a set of base pair displacements hence by one point $\{x_0 , a_m , b_m\}$ in the path configuration space. This establishes a biunivocal correspondence between the molecules ensemble and $N_{eff}$ while the overall base pair ensemble size is measured by $N_\tau \times N_{eff}$.

Applying the mapping technique in Eqs.~(\ref{eq:3}), (\ref{eq:5}) to Eq.~(\ref{eq:1}), the classical partition function for the DNA molecule in the solvent is written as

\begin{eqnarray}
Z_C=& &\oint \mathfrak{D}x\exp\bigl[- \beta A_C\{x\}\bigr]\, \nonumber
\\
A_C\{x\}=& & \sum_{i=\,1}^{N_\tau} \Bigl[{\mu \over 2}\dot{x}(\tau_i)^2 + V_S(x(\tau_i),x(\tau')) + \, \nonumber
\\
& &V_M(x(\tau_i)) + V_{sol}(x(\tau_i)) \Bigr] \, \nonumber
\\
\oint \mathfrak{D}x\equiv & &{1 \over {\sqrt{2}\lambda_\mu}}\int dx_0 \prod_{m=1}^{M_F}\Bigl({{m \pi} \over {\lambda_\mu}}\Bigr)^2 \int da_m \int db_m \, \, , \, \nonumber
\\
\label{eq:6c}
\end{eqnarray}

where  $\lambda_\mu$ is the thermal wavelength which, by virtue of Eq.~(\ref{eq:6}), takes the expression ${\lambda_\mu}=\,\sqrt{{\pi } / {\beta K}}$. $\mathfrak{D}x$ is the measure of integration and $\oint$ indicates that the paths $x(\tau)$ are closed trajectories \cite{io3,io5}.

\section*{IV. Thermodynamical Results}

The theory is applied to a fragment with $N_\tau =\,100$ base pairs whose sequence is:

\begin{eqnarray}
& &GC + 6AT +  GC + 13AT + 8GC + AT + 4GC + \nonumber
\\
& &AT + 4GC + AT + 8GC + [49-100]AT \, . \,\nonumber
\\
\label{eq:10}
\end{eqnarray}

Due to the prevalence of AT- base pairs, bubbles may open in the leftmost and, more likely, in the right part. The left portion of $48$ base pairs has the same sequence as the L48AS fragment of Ref.\cite{zocchi2} although our model cannot distinguish, for instance, a AT- followed by AT- along the backbone from AT- followed by TA-pair. Such differences may shift considerably the melting temperature in short fragments \cite{subirana,santa} with effects which are quantitatively not fully understood. Coherently with the notation used in Ref.\cite{io1}, the sequence in Eq.~(\ref{eq:10}) will be hereafter named {L48AT22}.

\subsection*{A. Fractions of Open Base Pairs}

The melting temperature is experimentally defined as the
temperature at which half of the molecules in the sample are in the double-helical
state and half are in the single strand, random-coil state. As explained in the previous Section, this amounts to say that half of the configurations $N_{eff}$ for the fragment in Eq.~(\ref{eq:10}) are closed and half are open.
When base pairs dissociate the UV signal changes quite abruptly  \cite{inman}. However, once the UV signal measures that half of the base pairs are open, this may indicate \emph{either} that half of the molecules in the sample are open and half are closed \emph{or} that all molecules are half-open. Accordingly absorption methods cannot distinguish intermediate states for a single molecule configuration which, instead, would be quite interesting in order to understand the nature of the melting transition. New techniques based on quenching of single strands are becoming available to trap intermediate states \cite{zocchi2}.

On the theoretical side, the problem arises to define when a configuration is open or closed. There is necessarily some arbitrariness intrinsic to the Hamiltonian model as this is expressed in terms of  base pair stretchings which are continuous variables. Accordingly
one may assume that a configuration is open when all $N_\tau$ base pairs are larger than a certain threshold $\zeta$ which however cannot be set univocally. For these reasons even large discrepancies regarding the choice of $\zeta$ are found in the literature \cite{ares,joy09}.

To tackle these questions I have computed the fraction of open base pairs $F_{op}$ for the system in Eq.~(\ref{eq:10}) taking different $\zeta$ and searching for a range of values which yield a description of the denaturation  qualitatively consistent with that provided by thermodynamical indicators such as the specific heat \cite{bresl}. This reasoning is inspired by the observation that the fraction of closed base pairs, $1 - F_{op}$, is a measure of the system internal energy hence $d F_{op} / dT$ is proportional to the specific heat. This holds for a homogeneous chain but also in heterogeneous DNA the specific heat is an indicator of the melting as it displays sharp peaks at the temperatures where various parts of the sequence open. Thus some correlation is expected between $F_{op}$ and specific heat plots versus temperature.

Using Eq.~(\ref{eq:6c}) the fraction of open base pairs is given, in terms of path integrals, by

\begin{eqnarray}
& &F_{op} =\, {1 \over {N_\tau}}\sum_{i=1}^{N_\tau} \vartheta\bigl(< x(\tau_i) > - \zeta \bigr) \,  \nonumber
\\
& &< x(\tau_i) >=\,Z_C^{-1}\oint \mathfrak{D}x x(\tau_i) \exp\bigl[- \beta A_C\{x\}\bigr] \,. \,
\label{eq:8}
\end{eqnarray}

The Heaviside function $\vartheta(\bullet)$ is justified by the sharp increases in the UV signal at the denaturation steps.
In fact, in Eq.~(\ref{eq:8}), $x(\tau_i)$ is averaged over the configuration ensemble and only $< x(\tau_i) >$ is directly confronted to $\zeta$. Then, to be rigorous, every configuration might have at least a base pair such that, $x(\tau_i) < \zeta$, (and therefore to be classified as \emph{closed}) whereas the average configuration might still be \emph{open} as, $< x(\tau_i) > - \zeta > 0$, for any $i$. With this caveat and aware that there exist alternative definition for $F_{op}$ \cite{pey3,io11} directly relating the base pair stretching to the opening threshold, Eq.~(\ref{eq:8}) is hereafter used as it better captures the multiple steps in the denaturation.

\begin{figure}
\includegraphics[height=7.0cm,angle=0]{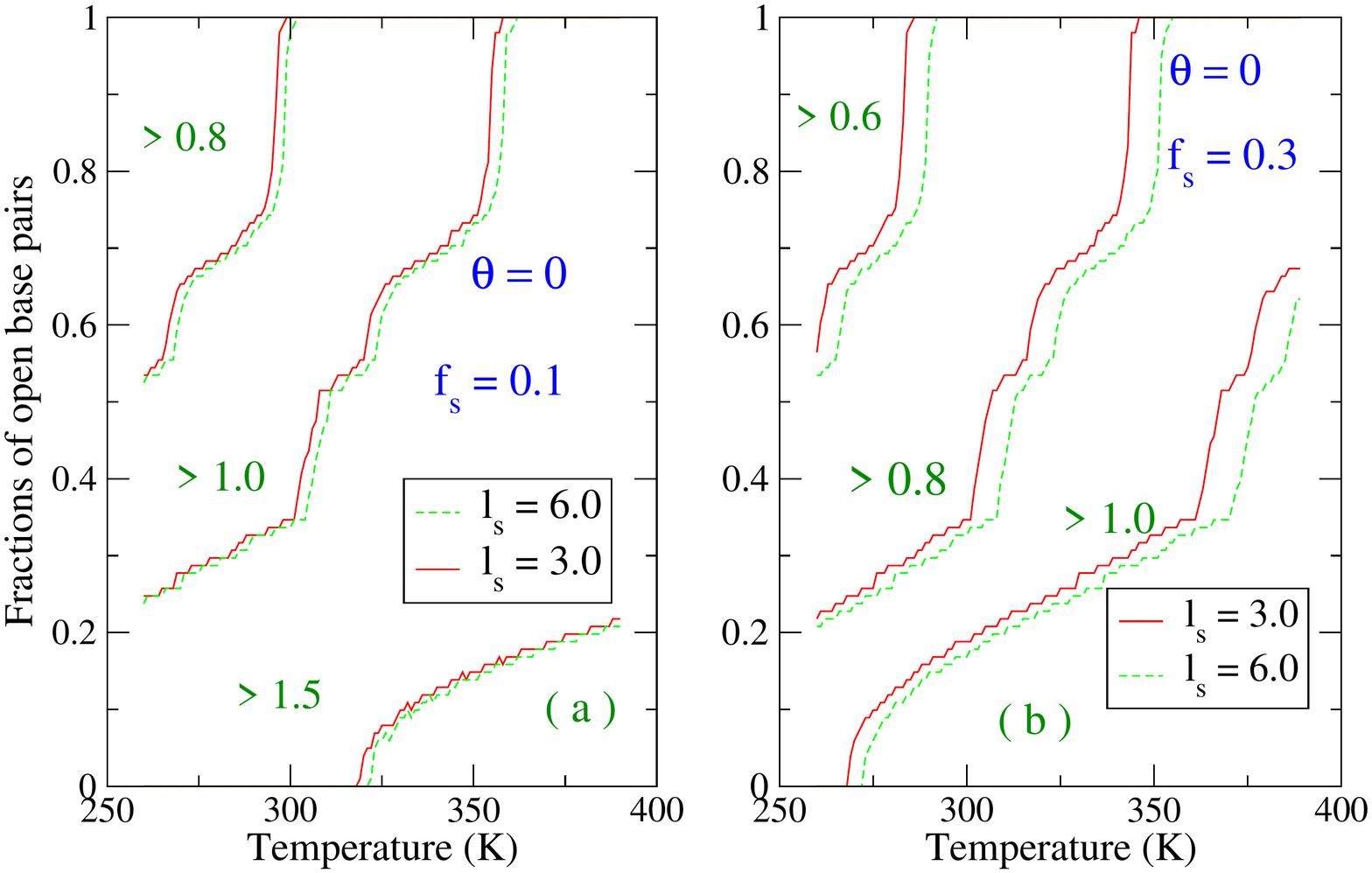}
\caption{\label{fig:2}(Color online) Fractions of open base pairs, calculated via Eq.~(\ref{eq:8}), for two solvent barrier factors:  (a) $f_s=\,0.1$, (b)   $f_s=\,0.3$. $l_s$ is in units ${\AA}$. Three thresholds $\zeta$ are assumed: (a) $\zeta =\,0.8, \, 1.0, \, 1.5 {\AA}$, \, (b) $\zeta =\,0.6, \, 0.8, \, 1.0 {\AA}$. The plots show the fractions of base pairs larger ($>$) than $\zeta$ versus temperature.}
\end{figure}

I have first evaluated the role of the solvent potential in Eq.~(\ref{eq:8}) assuming an untwisted DPB model, thereby $\theta=\,0$ in Eq.~(\ref{eq:1}). The results are reported in Fig.~\ref{fig:2} for those two barrier factors $f_s$ considered in Fig.~\ref{fig:1}. First one notices that $f_s$ has to be sufficiently large in order to make also the effect of $\l_s$ appreciable.  $F_{op}$ is calculated taking three $\zeta$  in Fig.~\ref{fig:2}(a) and Fig.~\ref{fig:2}(b) respectively. Two values, $\zeta=\,0.8, 1{\AA}$, are common to both figures. It is seen that the solvent has a stabilizing effect as, for the same $\zeta$ but larger $f_s$, $F_{op}$ attains the unity at higher temperatures. In Fig.~\ref{fig:2}(a), all averaged paths become larger than $\zeta=\,0.8 {\AA}$ at $T \sim 300K$ whereas this event occurs, in Fig.~\ref{fig:2}(b), at $T \sim 350K$: a sizeable shift which points to the importance of the solvent in modeling DNA fragments with the purpose to fit experiments. Certainly the fact that all average base pair paths are larger than, say $\zeta=\,0.8{\AA}$, at a given $T$ does not necessarily mean that the sequence is melting at that $T$. However, had we to know the melting temperature $T_m$ for a specific fragment, we would able to estimate via Eq.~(\ref{eq:8}) the $\zeta$ value such that $F_{op}=\,1$ at that $T_m$. In this view the computational method can select a reliable threshold for the base pair elongations with respect to the equilibrium. Above such threshold the average configuration may be considered as open. In substantial agreement with Refs.\cite{ares,pey3} our numerical work suggests that a range $\zeta \sim [0.5 - 1.0]{\AA}$ is appropriate to the purpose of modeling a multistep denaturation which occurs at $T_m \sim [300 - 390]K$. This $\zeta-$ range is chosen for the next calculations together with $f_s=\,0.3$ and $l_s=\,3.0 {\AA}$.

\begin{figure}
\includegraphics[height=7.0cm,angle=0]{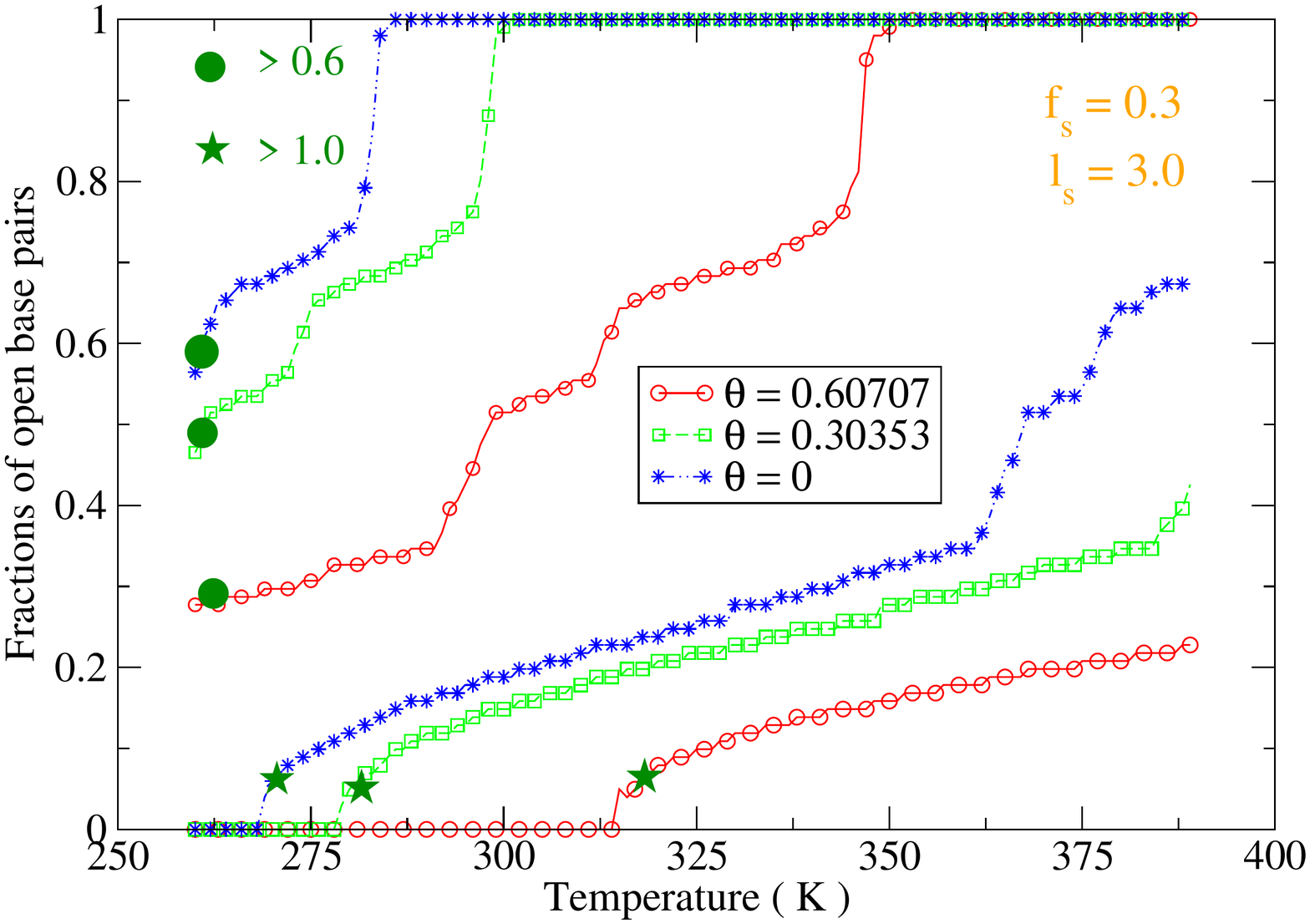}
\caption{\label{fig:3}(Color online)  Fractions of open base pairs versus temperature calculated according to Eq.~(\ref{eq:8}). The green filled circles and stars indicate the fractions of base pairs larger than the thresholds $\zeta=\,0.6 {\AA}$ and $\zeta=\,1.0 {\AA}$ respectively. Three twist angles $\theta$ (in \emph{rad}) are assumed in the stacking potential $V_S$ of Eq.~(\ref{eq:1}. The backbone coupling is $K=\,60 meV {\AA}^{-2}$. $l_s$ is in units ${\AA}$. }
\end{figure}

Next we turn to the core of the work by estimating the effect of the twist on Eq.~(\ref{eq:8}). Beside the untwisted case ($\theta=\,0$), two angles $\theta=\,0.60707, \, 0.30353 \,rad$ are considered which correspond to accommodate in one turn of the helix about $10$ and $20$ base pairs respectively. The results are plotted in Fig.~\ref{fig:3} where $F_{op}$ for the sequence L48AT22 is calculated versus temperature assuming two threshold values.

$F_{op}$ varies strongly with $\theta$ and gets larger in the untwisted model suggesting that the latter may denaturate at lower temperatures. Setting $\zeta=\,0.6{\AA}$ one observes that, by increasing $\theta$, $F_{op}=\,1$ is attained at $T\sim \,275, \,300, \, 350K$ respectively. This does not allow us to conclude that the case $\theta_{eq}=\,0.60707 \,rad$ corresponds to \emph{the stablest} configuration for the fragment but Fig.~\ref{fig:3} suggests that twist angles smaller than $\theta_{eq}$ would produce an helicoidal geometry which undergoes large fluctuational effects already below room temperature.

\begin{figure}
\includegraphics[height=7.0cm,angle=0]{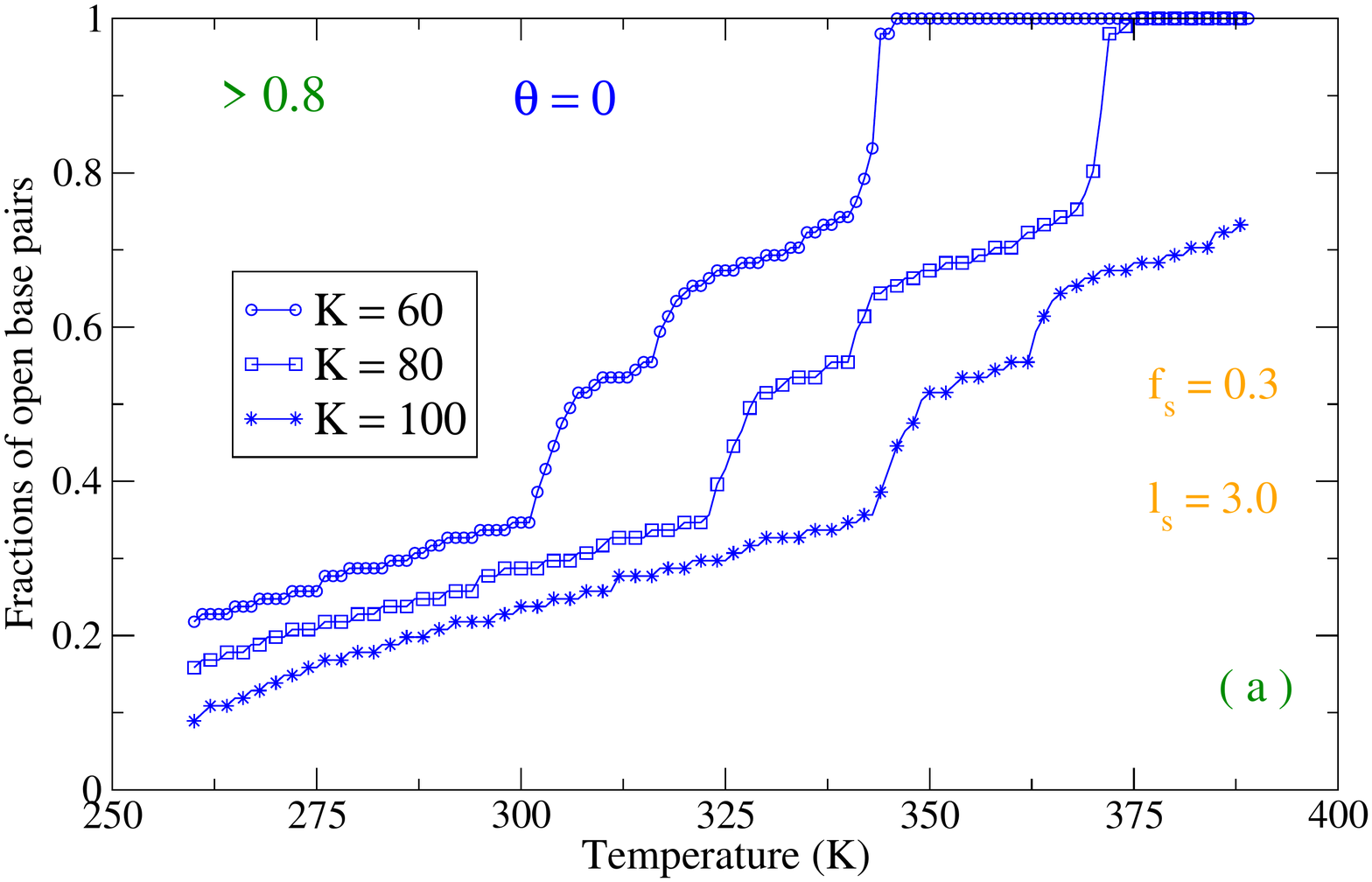}
\includegraphics[height=7.0cm,angle=0]{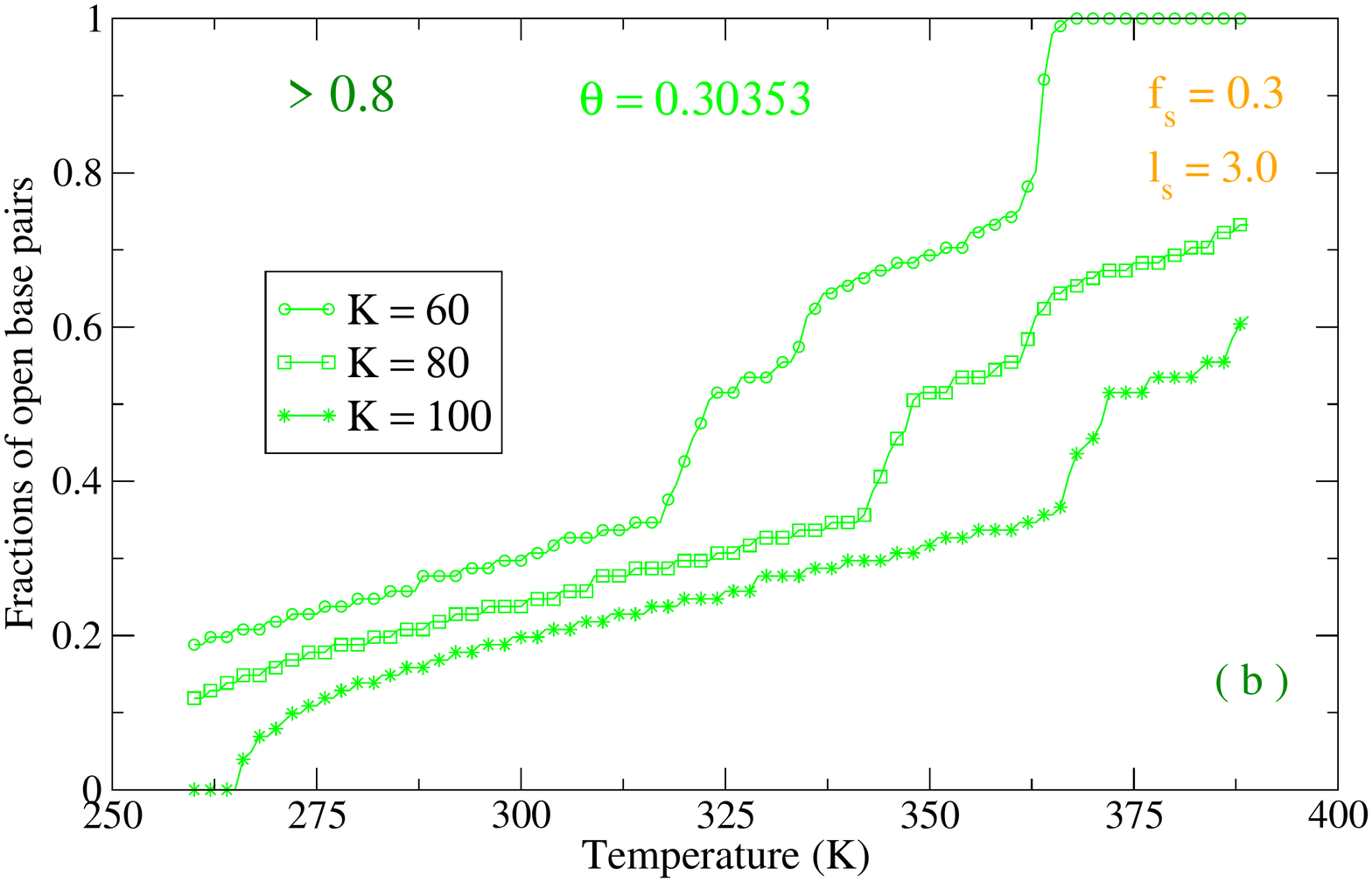}
\includegraphics[height=7.0cm,angle=0]{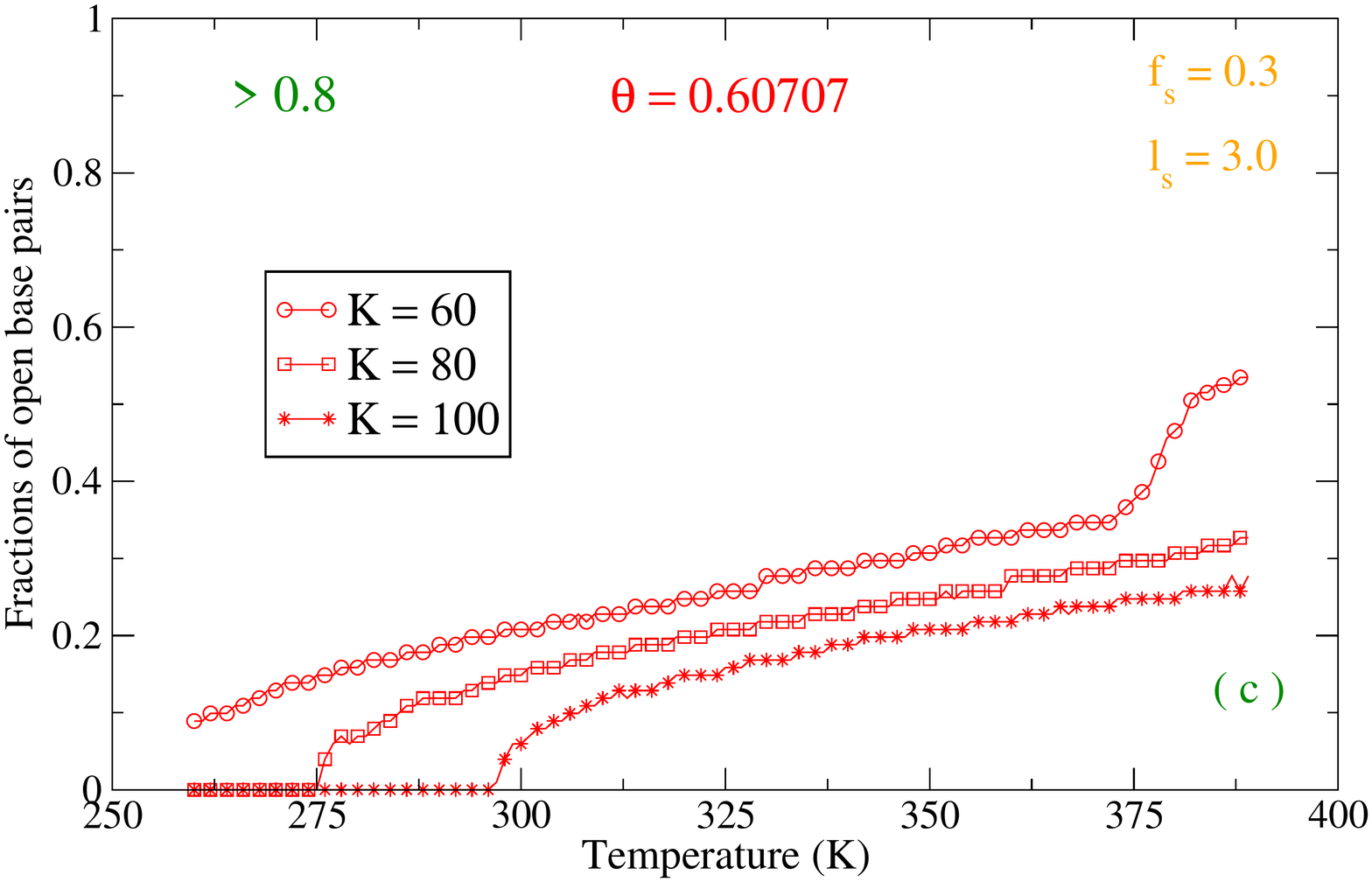}
\caption{\label{fig:3n}(Color online)  Fractions of open base pairs, larger than the thresholds $\zeta=\,0.8 {\AA}$, calculated through Eq.~(\ref{eq:8}) for different backbone couplings $K$ (in units of $meV {\AA}^{-2}$).  The three twist angles $\theta$ of Fig.~\ref{fig:3} are assumed in (a), (b) and (c) respectively. $l_s$ is in units ${\AA}$. }
\end{figure}

So far the melting profiles have been computed taking a constant harmonic stacking $K$. As the overall stability conditions for the double helix \cite{yakov} are determined by the base stacking it is worth analyzing $F_{op}$ as a function of $K$ in the presence of the solvent potential \cite{next}. The results are shown in Fig.~\ref{fig:3n} for the twists considered in Fig.~\ref{fig:3}. The threshold is set at $\zeta=\,0.8 {\AA}$ which corresponds to the maximum in the barrier height for GC-pairs due to the solvent (see Fig.~\ref{fig:1}). The strength of $K$ has a large weight in the untwisted structure (Fig.~\ref{fig:3n}(a)): for instance, focusing on $T \sim 340K$, a $K$ enhancement from $60$ to $80 meV {\AA}^{-2}$ produces a $\sim 40\,\%$ drop in $F_{op}$, whereas even $\sim 70\,\%$ of the average base pair displacements become smaller than $\zeta=\,0.8 {\AA}$ in the case $K=\,100 meV {\AA}^{-2}$. The effect of the backbone stiffness is much less dramatic in the twisted configurations as the twist on its own gives stability to the system. This can be observed in Fig.~\ref{fig:3n}(b) and, more remarkably, in Fig.~\ref{fig:3n}(c). In the latter, at $T \sim 340K$, $F_{op}$ drops by only $\sim 10\,\%$ in going from $60$ to $100 meV {\AA}^{-2}$. Note however that the melting feature at $T \sim 380K$ displayed by the $K=\,60$ plot is smeared at larger $K$ confirming that accurate determinations of the stacking parameters in heterogeneous sequences are fundamental for quantitative predictions of the melting profiles.  The backbone coupling also affects the degree of cooperativity in the denaturation transition as $N_{eff}$ is strongly reduced by increasing the stiffness via $K$ \cite{io11}.

\subsection*{B. Specific Heat versus Twist}

\begin{figure}
\includegraphics[height=7.0cm,angle=0]{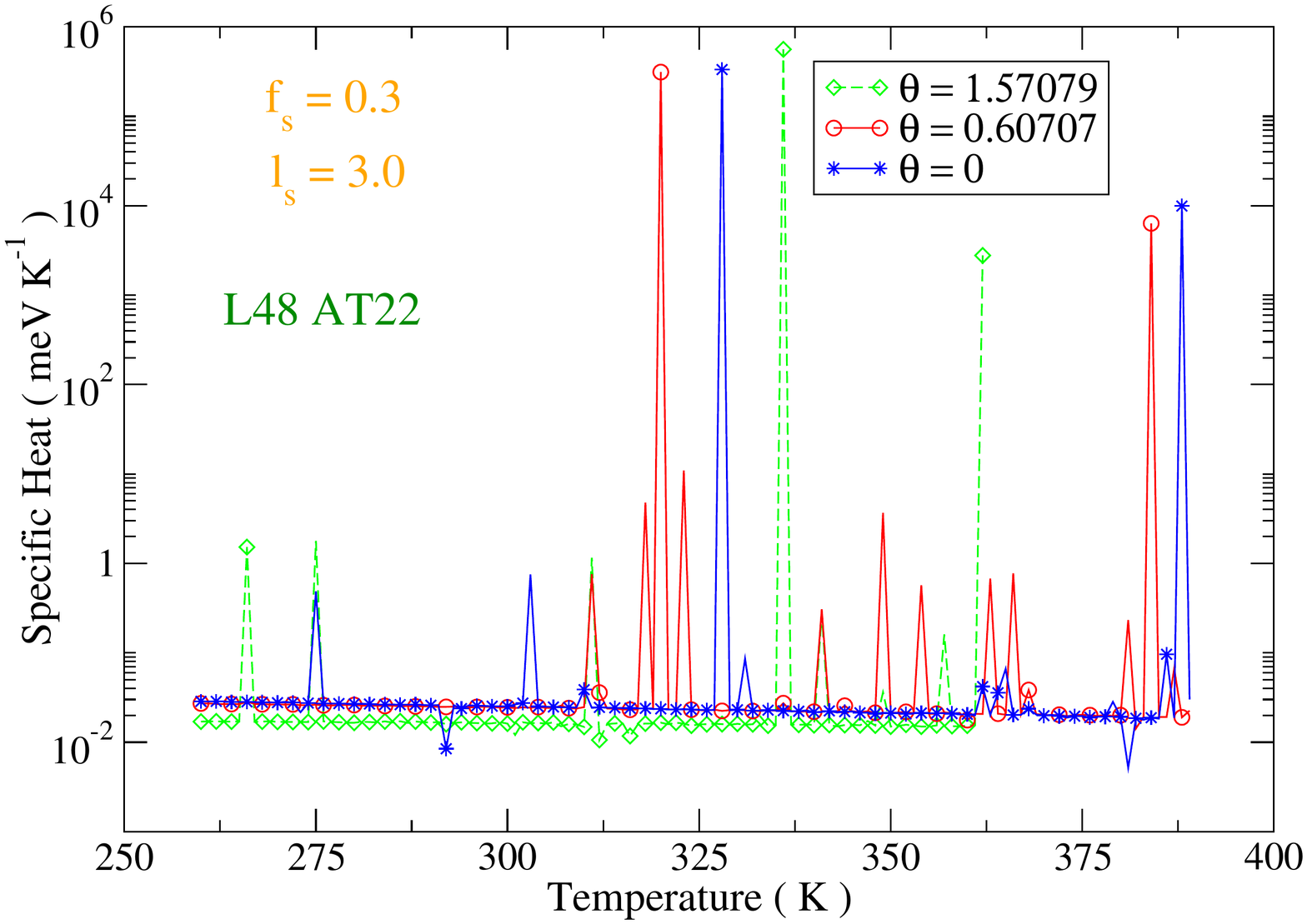}
\caption{\label{fig:4}(Color online)  Specific Heat versus temperature calculated via Eq.~(\ref{eq:6c}) by varying the twist angle $\theta$ (in \emph{rad}). The backbone coupling is $K=\,60 meV {\AA}^{-2}$.}
\end{figure}

Computation of Eq.~(\ref{eq:6c}) yields the free energy $A=\, -{\beta^{-1}}\ln Z_C$ hence the thermodynamic properties of the fragment in Eq.~(\ref{eq:10}).
For any choice of the model parameters the entropy is always found to be a continuously growing function of temperature confirming our previous analysis regarding the character of the denaturation: this is a smooth crossover which, in heterogeneous fragments, appears in multisteps driven by the AT-rich regions of the sequence. While the entropy plots (similar to those in Ref. \cite{io1}) are not displayed here, I focus on the specific heat which is shown in Fig.~\ref{fig:4} both for the untwisted case and two twist angles, $\theta=\,0.60707, \, 1.57079 \,rad$. The harmonic backbone coupling is set at $K=\,60 meV {\AA}^{-2}$.

The specific heat for $\theta=\,0.30353\,rad$  (the value taken in Fig.~\ref{fig:3}) is not reported as it would almost overlap with the untwisted curve. The plot with $\theta=\,0$ shows two shoulder peaks at $T \sim 275, \,300K$ which are consistent with the sharp increase in $F_{op}$, for $\zeta=\,0.6{\AA}$, below room temperature.  The main peak at $T \sim \,330K$ signals the denaturation of a large portion in the ensemble configuration and, confronting with Fig.~\ref{fig:2}(b), the corresponding opening threshold is $\zeta=\,0.8{\AA}$. However not all molecule configurations have opened at $T \sim \,330K$: some more do it at $T \sim \,388K$ as shown by the further shoulder peak witnessing that the solvent has effectively stabilized the system up to this value. In fact the $T \sim \,388K$- peak is smeared by switching off $f_s$.
Note that the two dips at $T \sim \,290,\,380K$ don't have any thermodynamical meaning: both can be smeared by slightly enhancing the path ensemble size rather indicating an overall lower stability for the ladder configuration of the untwisted DPB model.

Similar is the trend for the $\theta=\,1.57079\,rad$ specific heat presenting small size peaks below room temperature,  a main peak at $T \sim \,336K$ and a shoulder peak at $T \sim \,362K$. Instead, the $\theta_{eq}$- plot does not show any low $T$ peak indicating that the system is substantially stable up to room temperature. The main peak occurs at $T \sim \,320K$ where $\sim 65 \%$ of the base pair configurations have average values larger than $\zeta=\,0.6{\AA}$. All configurations exceed such threshold at $T \sim \,350K$ but this does not suffice to achieve complete melting as some base pairs open at higher $T$ once their stretchings exceed a larger $\zeta$ ($\sim \, 0.8{\AA}$). These findings are qualitatively consistent with a neutron scattering investigation of B-DNA \cite{theo2} showing the persistence of closed base pairs regions well inside the denaturation regime while the melting transition appears overall smooth.

The specific heat curves reveal multistep melting with fluctuational openings for all investigated twist angles. However  only the system with $\theta_{eq}$ does not present denaturation peaks below room temperature pointing to a higher stability for such helicity.  This result, in itself peculiar, is a motivation to analyze further the model dependence on the twist.

\subsection*{C. Base Pairs Path Ensemble versus Twist}

\begin{figure}
\includegraphics[height=7.0cm,angle=0]{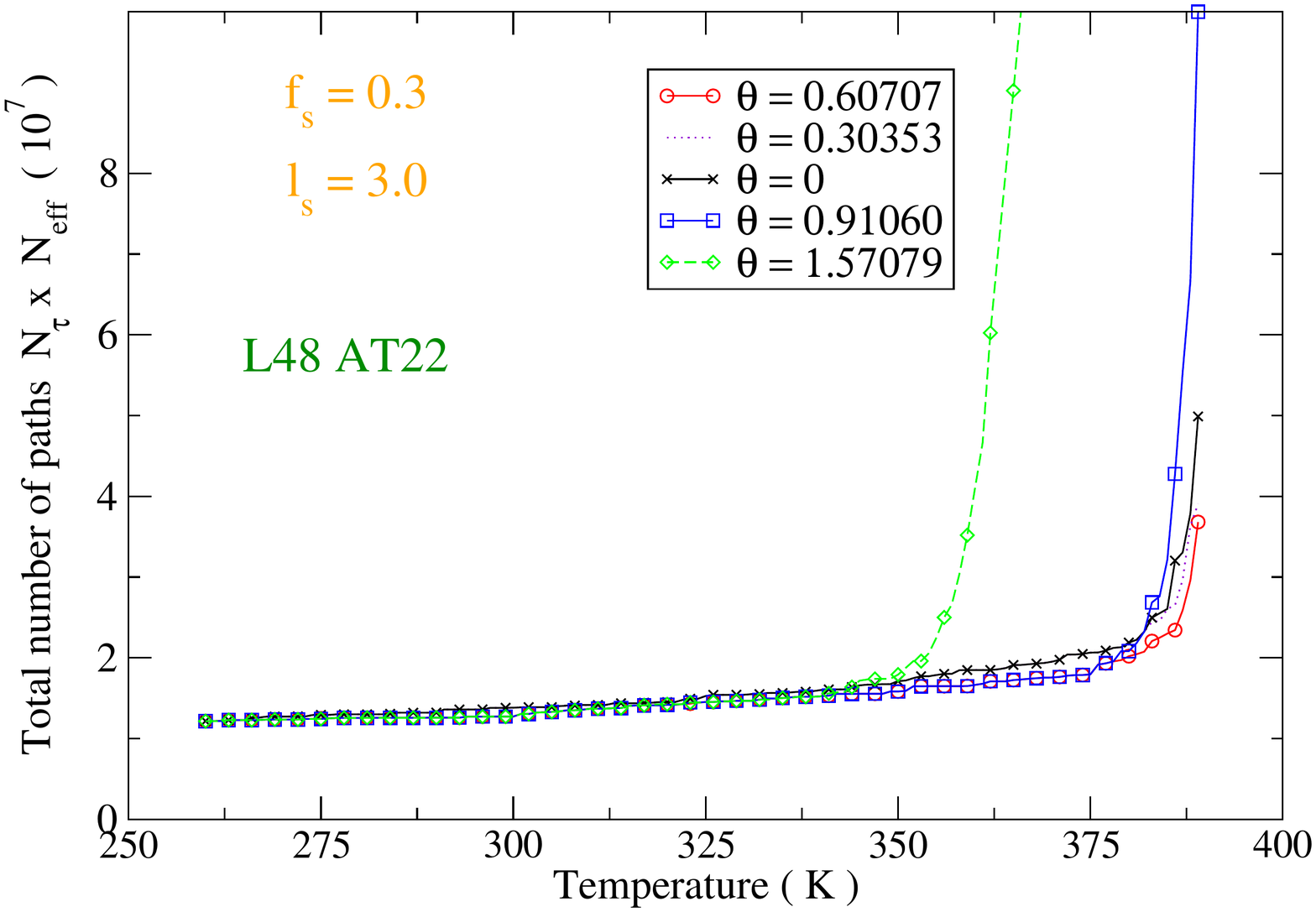}
\caption{\label{fig:5}(Color online) Total number of paths used in the computation of the DNA fragment thermodynamics for the untwisted case and four values of the twist angle $\theta$ (in \emph{rad}). The backbone coupling is $K=\,60 meV {\AA}^{-2}$.}
\end{figure}

To complete the study, I consider the size of the configurations ensemble for the base pairs in the fragment which is defined by $N_\tau \times N_{eff}$. It says how many distinct path configurations participate to $Z_C$ at any temperature thus providing a measure of the degree of cooperativity in the system. In homogeneous DNA, close to the melting transition, such degree grows sharply \cite{io} as many base pairs are involved in the event at about the same $T$. In heterogeneous DNA the cooperativity degree grows continuously as the transition is spread out in multisteps \cite{hwa}. Although the $T$ locations of the denaturation steps is not precisely signalled by $N_\tau \times N_{eff}$, the more stable is the heterogeneous fragment against denaturation the smaller $N_\tau \times N_{eff}$ is expected to be. In a way, this fundamental parameter for the path integral method measures the molecule resilience to the thermally induced perturbation.
In Fig.~\ref{fig:5}, $N_\tau \times N_{eff}$ is plotted versus $T$ for the untwisted case and four twist angles. In all cases the calculation starts at $T=\, 260K$ including  $\sim \,12 \times 10^6$ paths which ensure numerical convergence for $Z_C$. Then, $N_\tau \times N_{eff}$ is re-normalized upwards versus temperature according to the method described in Section III. As a main result, the helicity $\theta_{eq}$ is indeed the value providing more stability to the fragment. For all other considered twists $\theta$, $N_\tau \times N_{eff}$ grows faster versus $T$ indicating that the double strand molecule is more easily disordered by thermal effects on the hydrogen bonds. This means that denaturation steps may occur at lower $T$ than for the $\theta_{eq}$ case although no proportionality relation between $N_\tau \times N_{eff}$ and twist angle can be generally inferred.  Note in fact that the case $\theta=\,1.57079 \,rad$, which displays partial openings already at low $T$ (Fig.~\ref{fig:4}), also shows the most rapid growth in the size of the path ensemble.
To capture these features one has to perform summations over $\sim 10^8$ base pair paths in the upper portion of the considered $T$ range, a highly time consuming computational task.

The twist $\theta_{eq}$ is found to yield the highest stability also for other sequences obtained by varying order and relative content of AT- and GC- pairs in Eq.~(\ref{eq:10}).

This study has sampled different twist configurations searching for the most energetically advantageous. The mesoscopic Hamiltonian in Eq.~(\ref{eq:1}) does not account for torsionally constrained or enzyme driven supercoiling effects \cite{strick}. However, by virtue of the closed boundary conditions on the path displacements, it is worth to investigate whether the model may provide informations for the occurrence of supercoiling in small circular DNA molecules \cite{marko1}. Say $\sigma$ the superhelical density which measures the degree of supercoiling with respect to the relaxed molecule:
as natural DNA is negatively supercoiled, underwound configurations should be energetically more favored than the overwound ones \cite{calla}. Assuming zero number of writhes, $N_\tau \times N_{eff}$ is computed by varying the number of twists $Tw$ with respect to the equilibrium, $(Tw)_0=\,10$. This allows to monitor the cooperativity degree tuning the twist angle $\theta$ around $\theta_{eq}$. For $|Tw - (Tw)_0| =\,\mathbb{Z}$ with integer $\mathbb{Z}$, one gets two angles $\theta_{i=\,1,2}$. E.g., $\mathbb{Z}=\,1$ implies:\emph{ a)} $Tw=\,11$ and $\theta_1=\,0.69115$ hence, $\sigma > 0$; \emph{b)} $Tw=\,9$ and $\theta_2=\,0.56548$ hence, $\sigma < 0$. As $N_\tau =\,100$ for the sequence L48AT22, the helical configurations obtained by taking $\mathbb{Z} \in [1,3]$ will suffice to our purposes.

With the potential parameters of Fig.~\ref{fig:5}, the model indeed accounts for the asymmetry between positive and negative supercoiling proving the energetic convenience of the $\sigma < 0$ configurations.

In fact, it is found that:

\begin{eqnarray}
& &\mathbb{Z} =\,1 \,  \nonumber
\\
& &N_{eff}(\theta_1) =\, N_{eff}(\theta_2) =\,N_{eff}(\theta_{eq})\,  \nonumber
\\
\label{eq:19}
\end{eqnarray}

throughout the whole temperature range. Instead:

\begin{eqnarray}
& &\mathbb{Z}=\,2 \,,3  \nonumber
\\
& &N_{eff}(\theta_1 > \theta_{eq}) > N_{eff}(\theta_2 < \theta_{eq}) \sim \,N_{eff}(\theta_{eq})\,  \nonumber
\\
\label{eq:20}
\end{eqnarray}

in the upper portion of the temperature range.

This amounts to say that small twist changes, i.e. $Tw=\,9,\,11$,  do not perturb the thermodynamics of the relaxed molecule but once $\theta_{i=\,1,2}$ differ significantly from $\theta_{eq}$, i.e. for $Tw=\,8,\,12$,  it is the underwound configuration to be naturally preferred: a moderate helix unwinding can sustain local denaturation bubbles which absorb the released twisting strain thus allowing for an overall molecule stability. Certainly, a too large unwinding ($\sigma \rightarrow -1$) favors the complete strand separation and leads to melting.

These findings are consistent with careful single molecule micromanipulation experiments \cite{strick} and theoretical predictions based on the Poland-Scheraga model \cite{metz10}, mostly in regard to the asymmetric system dependence on the sign of $\sigma$. At this stage however our method, applied to the sequence in Eq.~(\ref{eq:10}), does not display any relevant physical effect by varying the superhelical density in the biologically interesting range, $|\sigma| < 0.1$.

\section*{V. Concluding Remarks}

The thermodynamics of a short fragment of heterogeneous DNA has been studied by the path integral method focusing on the temperature range in which denaturation takes place. The model Hamiltonian contains a solvent interaction term that enhances the base pair dissociation energy and stabilizes the hydrogen bonds between complementary strands. Overcoming the limitations intrinsic to previous investigations which model the double helix through two parallel chains, I have introduced as a free parameter a twist angle between two adjacent base pair stacked along the molecule backbone. In B-DNA the strands twist, around the molecule axis, about once every ten base pairs hence, the equilibrium twist angle is $\theta_{eq}\simeq \,0.60707 \,rad$. Assuming an \emph{adenine-thymine} rich sequence of one hundred base pairs, I have studied whether the path integral model can capture an energetic advantage for such equilibrium geometry. The base pair displacements are considered as imaginary time dependent paths and the ensemble of \emph{good paths} which contribute to the system partition function has been built consistently with some physical constraints, markedly the shape of the hydrogen bond potential and the second law of thermodynamics. Such path ensemble is a temperature dependent representation of the possible molecule configurations and it accounts for those fluctuational effects which are key to the DNA dynamics, mainly in short fragments.  In general, close to the denaturation, the path ensemble size grows as the system becomes more cooperative and more paths participate to the transition. Thus, the ensemble size measures the overall system stability against thermal disruptions of the hydrogen bonds. By varying the twist angle, it is found that the stability is higher precisely for the torsion defined by $\theta_{eq}$ and this conclusion does not depend on the specific sequence. The Ansatz made in the Introduction is then proved in the light of the path integral approach here developed.

Two major physical effects appear in our model and compete on the energy scale: the base pair displacements should become large enough to yield those fluctuational openings which are peculiar of the double helix dynamics and, at the same time, too large base pair fluctuations should be discouraged as they destabilize the system already at room temperature. The  stacking potential has the capability to select the appropriate fluctuation amplitudes as a function of the twist.

In \emph{untwisted or small twist models}, $ \theta \ll \theta_{eq}$, even large fluctuations are possible as the corresponding stacking energies, $V_S(y_n, y_{n-1})$, are low on the energy scale set by the one coordinate potential $V_M(y_n) + V_{sol}(y_n)$. Hence, such large fluctuations do contribute to partition function and thermal properties. It follows that molecules with small twist have scarce stability and begin to denaturate below room temperature.

On the other hand, in \emph{large twist models} with $ \theta \gg \theta_{eq}$, even small fluctuations have stacking energies which are higher than the hydrogen bonds dissociation energy hence, their contributions to the partition function become vanishingly small. Accordingly large twist models have scarce flexibility and do not account for those fluctuational openings that are vital to the molecule. In this view $\theta_{eq}$ emerges from the path integral computation as the energetically most convenient twist for the double strand configuration since it provides a right balance among the competing tendencies of this complex system.

For $ |\theta - \theta_{eq}| \sim 0.15$, the unwound double helix displays an energetic advantage over the overwound one indicating that the opening of local denaturation bubbles is a natural strategy to stabilize the system absorbing torsional strain.

I have also computed the fractions of open base pairs with reasonable opening thresholds and the specific heat for various twist angles: all the obtained results provide as much independent as consistent indications that the helicoidal geometry of B-DNA has indeed a higher stability, albeit maintaining that flexibility associated with the nonlinear character of the intra- and inter-strand interactions. Further investigations on these issues are due to come. In particular the twist should be treated as a rotational degree of freedom in a two dimensional path integral description thus incorporating also torsional fluctuation effects around the equilibrium geometry.


\begin{references}
\bibitem{next}
In fact, stacking heterogeneity should be included in the Hamiltonian model to account for the different opening probabilities of AT- and GC-base pairs: see \cite{krueg}.
\end{references}
\end{document}